\documentclass[a4paper, 10pt, conference]{IEEEtran}  
\usepackage{graphicx}
\usepackage[cmex10]{amsmath}
\usepackage[ruled,vlined,linesnumbered]{algorithm2e}
\usepackage{setspace}
\SetAlFnt{\small\sffamily}
\usepackage{caption}
\usepackage{subcaption}
\usepackage[simplified]{pgf-umlcd}
\usepackage{placeins}
\usepackage{booktabs}
\usepackage{hyperref}
\usepackage{textcomp}

\usepackage{amssymb}
\usepackage[hang,flushmargin]{footmisc}
\usepackage{lipsum}

\SetKwComment{Comment}{$\triangleright$\ }{}

\IEEEoverridecommandlockouts                              

\title{\LARGE \bfseries
Data-driven Thermal Anomaly Detection for Batteries using Unsupervised Shape Clustering
}

\author{
Xiaojun Li\textsuperscript{*}, Jianwei Li, Ali Abdollahi and Trevor Jones
\thanks{The Authors are with Gotion Inc, 48660 Kato Rd, Fremont, CA, 94538 USA.}
\thanks{Contact:\tt\small t.li@gotion.com}
}


\begin{document}

\maketitle

\IEEEpubidadjcol


\begin{abstract}

For electric vehicles (EV) and energy storage (ES) batteries, thermal runaway is a critical issue as it can lead to uncontrollable fires or even explosions. Thermal anomaly detection can identify problematic battery packs that may eventually undergo thermal runaway.  However, there are common challenges like data unavailability, environment and configuration variations, and battery aging. We propose a data-driven method to detect battery thermal anomaly based on comparing shape-similarity between thermal measurements. Based on their shapes, the measurements are continuously being grouped into different clusters. Anomaly is detected by monitoring deviations within the clusters. Unlike model-based or other data-driven methods, the proposed method is robust to data loss and requires minimal reference data for different pack configurations.  As the initial experimental results show, the method not only can be more accurate than the onboard BMS and but also can detect unforeseen anomalies at the early stage.

\end{abstract}



\section{Introduction}
\subsection{Background\label{challenge}}
Conventional anomaly detection methods for batteries usually depend on thresholds or lookup tables, often obtained by lab-testing of sample batteries, and may not apply to the individual battery that operates under different conditions. On the other hand, advanced anomaly detection methods, such as machine learning-based algorithms, require significant computational resources. In addition, traditional battery management systems (BMS) deployed on electric vehicles or energy storage systems are based on embedded micro-controllers, which lack computing power or memory to execute these algorithms on-board. 
\par
For a cloud-based BMS, the gathered data is transmitted to a data center for further analysis, during which advanced algorithms can be utilized. As a part of the fault diagnosis process, anomaly/fault detection is the most critical step. Based on timely detection, the BMS or vehicle controller (VCU) of the EV can take proper actions, which prevents a relatively small issue from developing into a severe problem. There are three mainstream methods for battery fault/anomaly detection: knowledge-based, model-based, and data-driven \cite{Hu2020}. The threshold-based method, which is the industry's standard practice, can be categorized as knowledge-based. In general, every battery manufacturer has its own heuristic "recipe" combined with testing data and engineering known-hows, which can not be generalized easily. Model-based approaches often include a physical model and an estimator \cite{plett2015battery}. In \cite{Lin2013OnlineMonitoring}, a lumped battery thermal model with time-variant internal resistance, altogether with an adaptive observer, are used to estimate the core battery temperature. \cite{Gao2019Micro-short-circuitModel} uses a cell difference model with an extended Kalman filter to estimate the micro-shot-circuit current, which can be used for short circuit diagnosis.  Some of the data-driven approaches depend on the recurrent neural network \cite{Ojo2020ABatteries,Li2021}. For example, \cite{Ojo2020ABatteries} uses a long short-term memory neural network to create residual signals for battery surface temperatures. A thermal fault is set when the residual is over a certain threshold. For model-based approaches, training data is needed for finding the optimal model parameters for different types of batteries, making it difficult and time-consuming to be implemented. Besides, data anomalies like long-time unavailability and shifting can cause the model to malfunction. Relatively speaking, training recurrent neural networks requires fewer efforts. However, the neural network also depends on the continuous signal influx. For both methods, battery degradation over time poses a challenge. In conclusion, the following common issues still pose significant challenges for battery anomaly detection. 

\begin{figure}[!t]
\centering
\includegraphics[width=1.05\columnwidth]{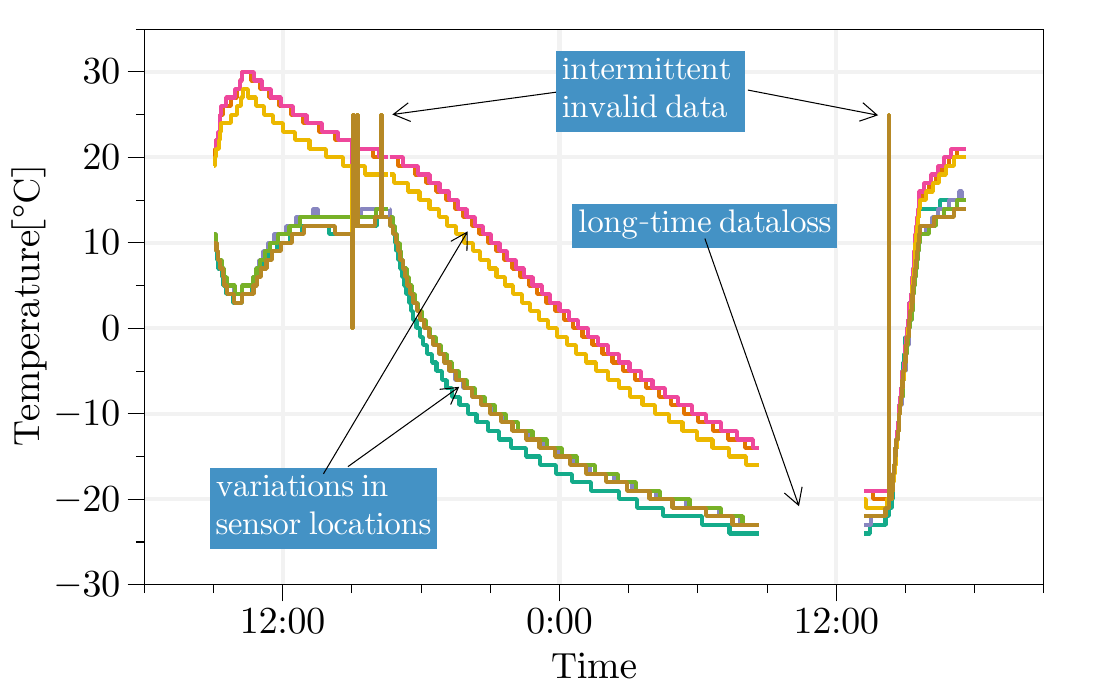}
\caption{Typical issues of thermal measurements for cloud-BMS data. In this figure, each signal represents the measurement from one of the temperature sensors installed on the battery pack. Data is from a fleet EV battery.}
\label{data_issue}
\end{figure}

\begin{itemize}
    \item \textit{Loss of data/invalid data}: For cost-saving, some wireless transmission modules in BMS or vehicles usually use less reliable networks such as 3G or even 2G. Due to network issues, loss of data and invalid data are common. As depicted in  Fig.~\ref{data_issue}, it may happen just intermittently or for an extended period. Data is also not available when the vehicle is shut down. Model-based approaches, such as Kalman filters and recurrent-neural networks, have internal states that rely on continuous data input. They may tolerate intermittent data loss, but long-term data loss will cause the model to lose track and give inaccurate results.
    \item \textit{Asynchronous data}: Because of communication or sensor issues, one or several signals may be sampled with some time delay. Residual-based methods, which evaluate all signals simultaneously, are likely to generate false positives with asynchronous data.
    \item \textit{Battery aging}: As the battery ages and deteriorates, the thermal and cell voltage measurement will deviate from their nominal range. Both residual and model-based method needs to adjust the threshold and parameters for them to work correctly. However, estimating the battery's state of health (SOH) itself is a challenging task \cite{Nuhic2013}. 
    \item \textit{Environment variations}: The environment variations, such as different geo-locations, sensor locations, and seasonal temperature changes, cause similar difficulties as battery aging. But they are harder to be factored in the model.
    \item \textit{Lack of training data}:  Electric vehicle and energy storage systems have different configurations. It is challenging to train the model so that it can adequately identify anomalies for all configurations. A tailored model is more accurate but is challenging to implement. Besides, labeled data may not be available for some configurations.
\end{itemize}

\subsection{The proposed method}

To address some of the challenges mentioned above, this paper proposes a data-driven, model-free approach that monitors the shape-similarities across the measurements to detect battery thermal anomalies. The proposed method does not require the data to be continuous, making it robust to data loss and invalid data. The same reason also alleviates the effects of battery aging or environment variations since cells are likely to deteriorate or be influenced by these variations as a whole. Asynchronous data issue is handled by the shape-based distance measurement, which is invariant to signal shifting. Furthermore, this method can be applied to different configurations easily and does not require any knowledge about sensor placements since it learns from the data. It also needs very little reference data (only several minutes).

\par

Clustering is an unsupervised learning algorithm that does not require training with labeled data. Therefore, this method can detect unforeseen anomalies. Furthermore, unlike the existing cross-similarity approaches\cite{Schmid2020}, the proposed method does not need feature extraction. Furthermore, the key advantage of this new method is the potential to give early warnings. By comparing shapes (which is scaling-invariant), this method can detect small signal deviations even if the value differences between measurements are still small, giving it the ability to capture abnormality at the early stage.

\section{The Anomaly Detection Method}

\subsection{The K-shape Clustering Algorithm}
Proposed by \cite{Paparrizos2015}, the K-shape clustering algorithm is intended to be used for time-series analysis. It adopts a cross-correlation sequence \(CC_\omega\left(\mathbf{x},\mathbf{y}\right)\) to determine signal similarity. The shape-based distance function is given by 

\begin{equation}
    SBD(\vec{x},\vec{y})\ =\ 1- \underset{\omega}{max}{\left(\frac{CC_\omega\left(\mathbf{x},\mathbf{y}\right)}{\sqrt{R_0\left(\mathbf{x},\mathbf{x}\right)R_0\left(\mathbf{y},\mathbf{y}\right)}}\right)}
    \label{k-shape}
\end{equation}

where \(\vec{x},\vec{y}\) are the normalized time-series measurements, and  \(R_0\) is the Rayleigh Quotient. The overall K-shape algorithm has two steps per iteration, repeated until convergence or max iteration is reached. In the first step ("assignment"), measurements are assigned to each cluster based on their similarity to the centroid. In the second step ("update"), the clusters’ centroids are updated. The overall time complexity $\mathcal{O}\left(max\left\{(ncm)log(m),nm^2,cm^3\right\}\right)$ scales linearly with the number of measurements $n$ and the number of clusters $c$ . This is a key feature that makes the proposed method applicable to large fleet battery systems or energy storage systems. The overall time complexity increases noticeably with the number of time-steps $m$. K-shape has been applied to time-series analysis and forecast \cite{Bega2019DeepCog:Learning,Gianniou2018Clustering-basedData,Calikus2019AHeating}, including battery cell voltage monitoring \cite{Haider2020}.

\subsection{Assumptions and Limitations}
Apart from the advantages given above, the proposed method relies on some assumptions and has its limitations.

    \subsubsection{The battery pack initially operates normally} We use the original operational data to acquire the reference cluster membership. It is worth mentioning that the proposed method only needs a small size of measurement data to extract the information. For example, in Section~\ref{exp}, only 4 minutes of normal operation data is used.
    \subsubsection{The anomaly does not occur in all measurements in the same manner} In theory, anomalies can cause no deviation in the measurement clustering if they affect all measurements simultaneously and in the same way, without disrupting the existing cluster memberships. Under such circumstances, the proposed method would not be effective. However, it is unlikely to happen in real life. A typical anomaly first appears only in one or a few measurements and can be detected by the proposed method.

\subsection{Methodology}

\begin{figure}[!t]
\centering
\includegraphics[width=2.5in]{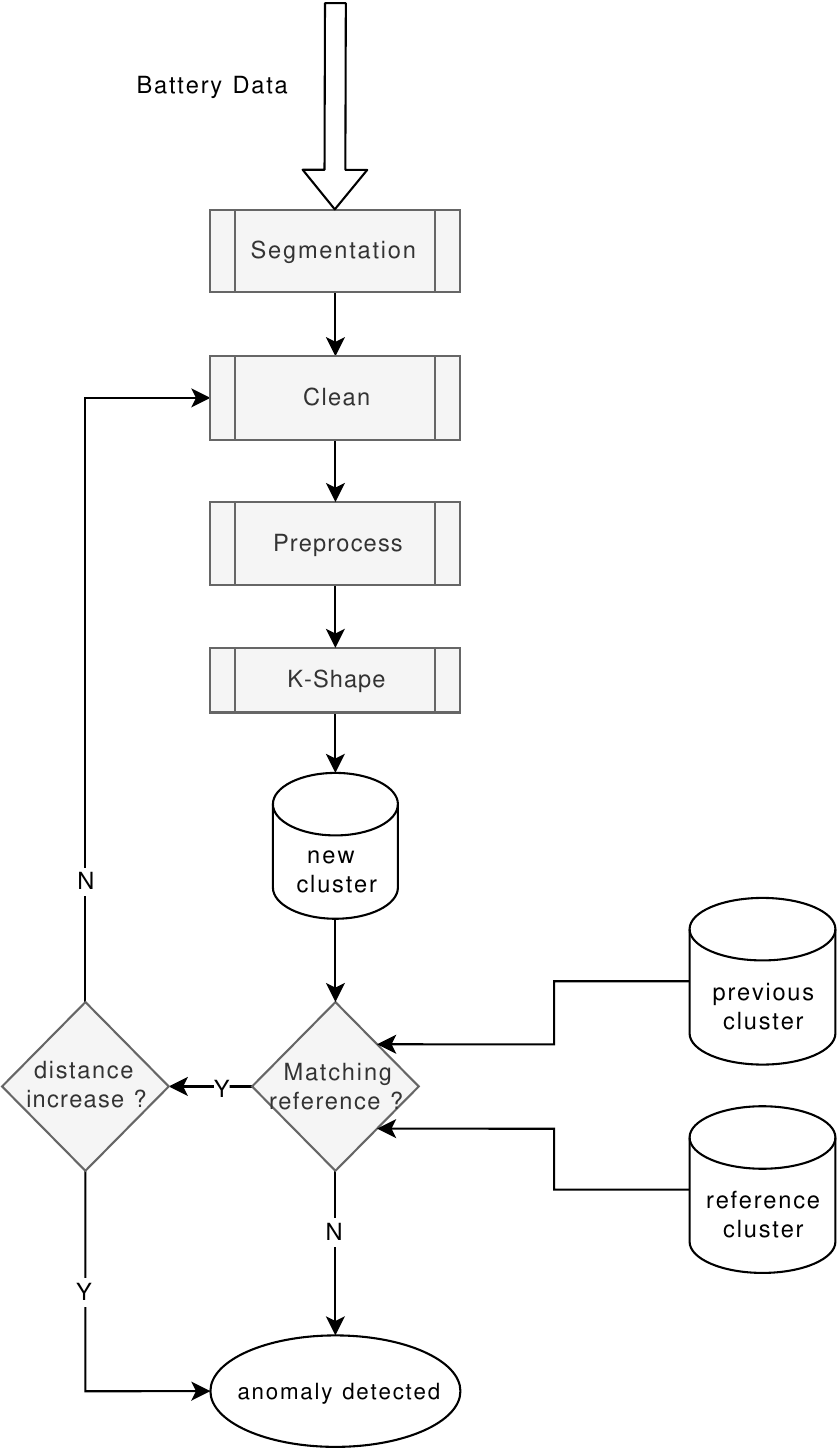}
\caption{Workflow of the proposed anomaly detection method}
\label{fig_sim}
\end{figure}

\begin{algorithm}[t]
\setstretch{1.25}
    \caption{${a[k],p[k] = AnmlyChk(\mathbf{mem}[k],\mathbf{dist}[k])}$}
    \algorithmfootnote{\begin{enumerate}
        \item $mem[0][i]$ denotes the original reference membership for the $ith$ measurement
        \item $\alpha$ denotes the weighting factor between long-term deviation and short-term change.
        \item An anomaly should be flagged when $p[k]$ is true or $d[k]$ is large.
    \end{enumerate}
         }
    \label{alg}
    \SetAlgoLined
    \KwIn{membership ($\mathbf{mem}[k]$) and distance ($\mathbf{dist}[k]$) for the $kth$ segment, which has $n$ measurements and $c$ clusters.}
    \KwOut{anomaly indicator ($p[k]$) and confidence level ($d[k]$) for the $kth$ segments}
    \tcp{initialization}
    $a[k] = False$ \\
    $p[k] = 0.$\\
    \tcp{membership change check}
    \For{$mem[k][i]$ in $\mathbf{mem}[k]$}
    {\uIf{$mem[k][i] \neq mem[0][i]$}
    {$a[k] = True $\\
    \KwRet{$(a[k],p[k])$}}}
    \tcp{distance change check}
     $dist[k] = \sum{\mathbf{dist}[k]}$\\
     \uIf{$dist[k-1] > \epsilon$  \tcp{avoid zero-division}}
     {$p[k] += (1 - \alpha)\cdot{sat}\left(\frac{{dist}[k]}{dist[k-1]}\right) +
     \alpha\cdot{sat}\left(\frac{{dist}[k]}{{dist}[0]}\right) $}
     \uElse{$p[k] +=  sat\left(\frac{{dist}[k]}{{dist}[0]}\right)$}
    
     \KwRet{$a[k],p[k]$}
\end{algorithm}

\begin{figure*}[ht!]
\centering
         \begin{subfigure}[b]{0.7\textwidth}
                 \centering
                 \includegraphics[width=\textwidth]{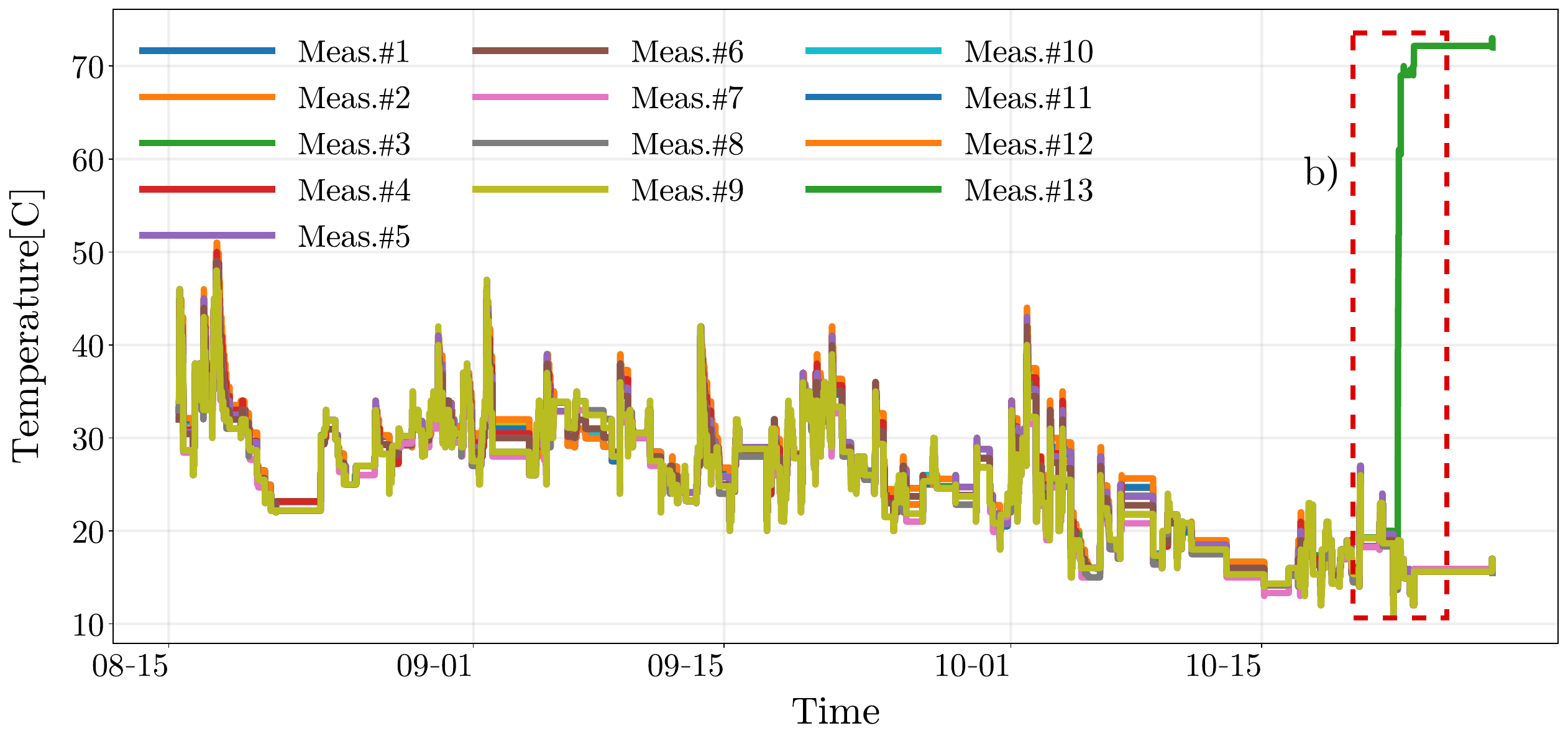}
                 \caption{}
                 \label{fig2:1-1}
         \end{subfigure}
     \begin{subfigure}[b]{0.35\textwidth}
             \centering
             \includegraphics[width=\textwidth]{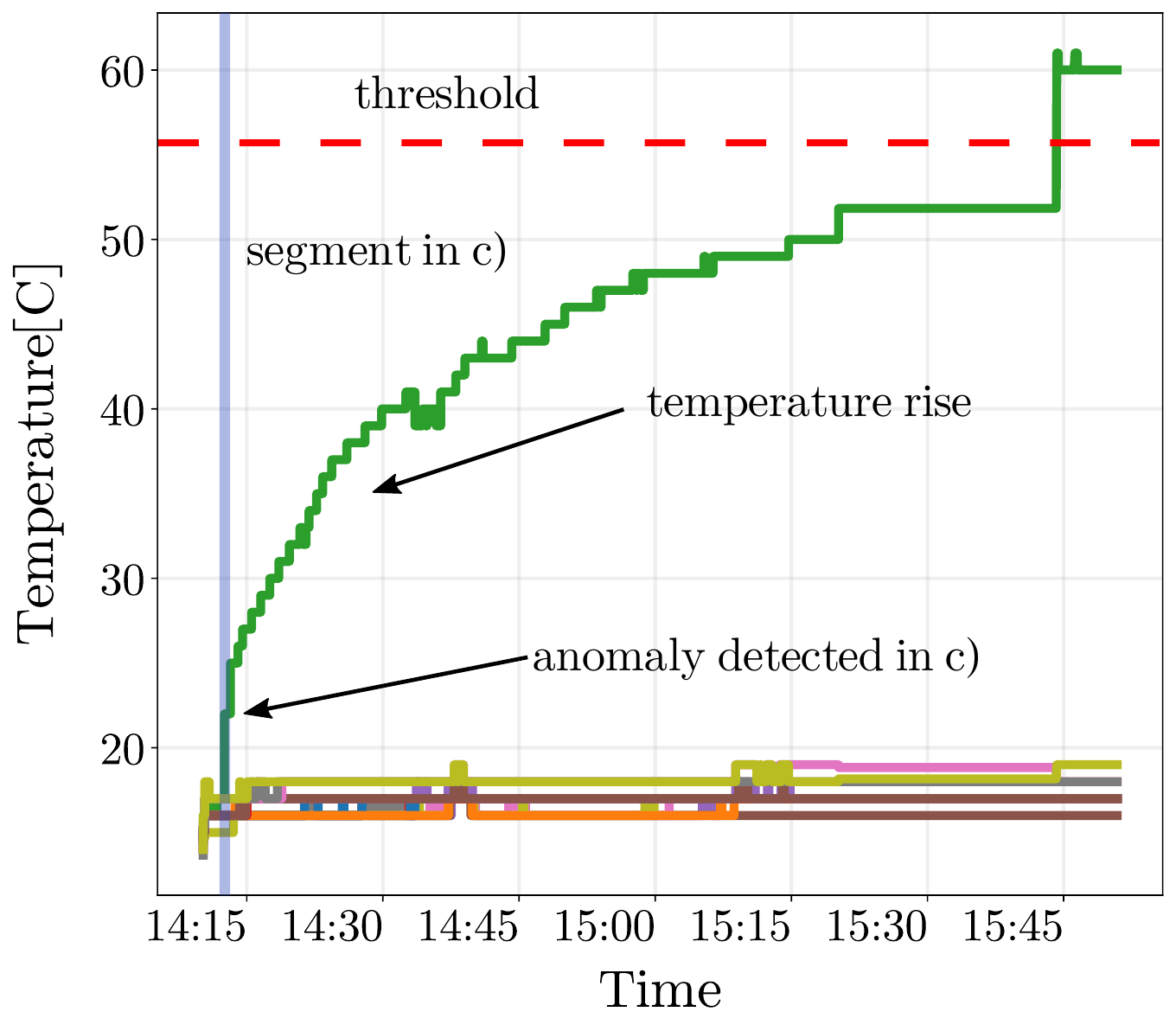}
             \caption{}
             \label{fig2:2-1}
     \end{subfigure}
     \begin{subfigure}[b]{0.35\textwidth}
             \centering
             \includegraphics[width=\textwidth]{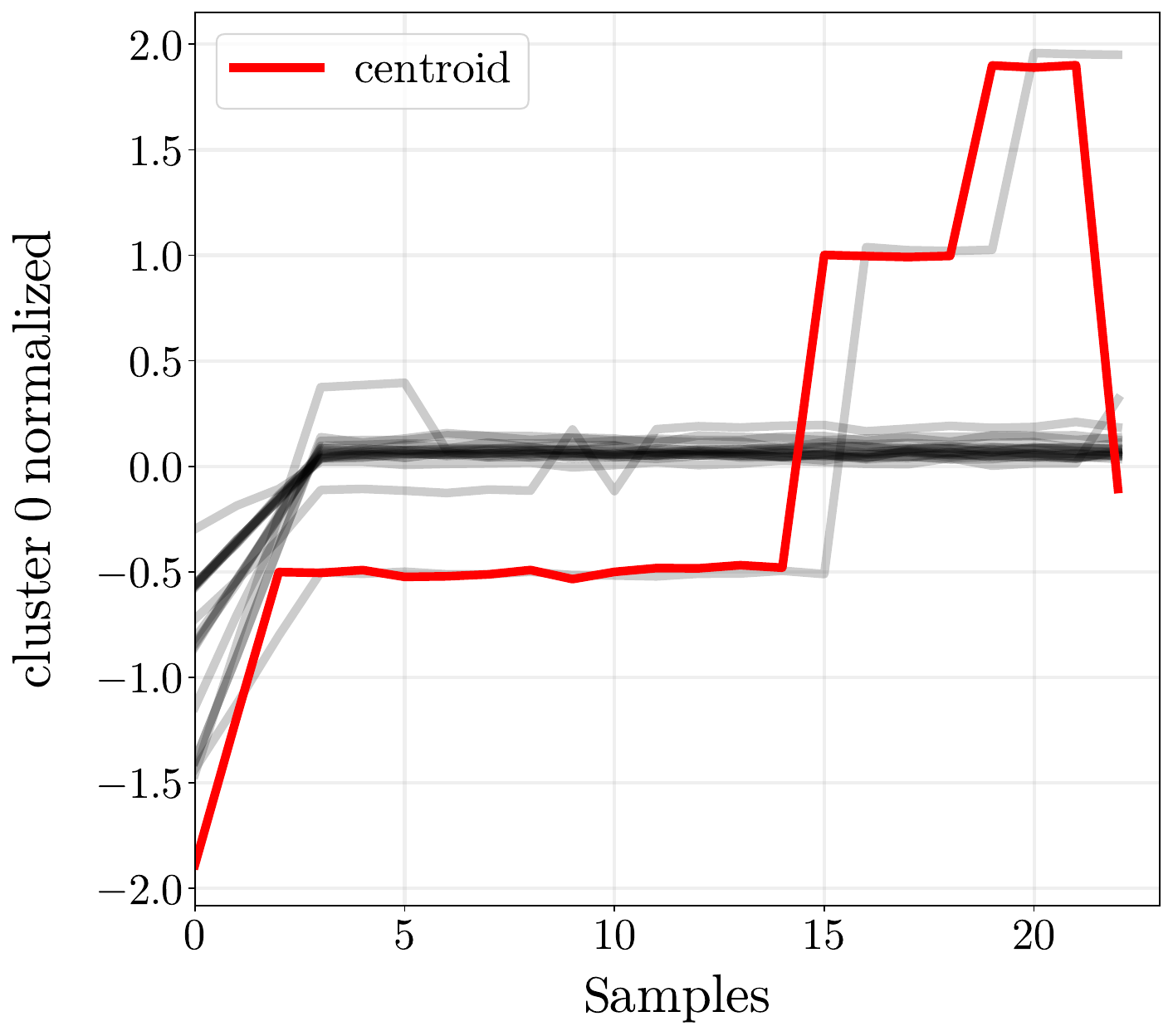}
             \caption{}
             \label{fig2:2-2}
      \end{subfigure}
\caption{Case I: (a) Thermal measurement data with the over-temperature fault. 
(b) The zoom-in view of fault occurrence, notice a temperature anomaly is detected by the proposed method 
90min before it surpasses the threshold. (c) The plot of measurement shapes for the segment (highlighted with light blue in (b)) where the anomaly is detected.} \label{fig2}
\end{figure*}

As the flowchart depicted in Fig.~\ref{fig_sim} , the proposed anomaly detection method contains the following stages:
\subsubsection{Segmentation/buffering} For offline implementation, existing data is segmented into smaller pieces, each of which will be clustered accordingly. For online implementation, data should be continuously collected and buffered as a segment, which is then processed. Because the K-shape's time complexity grows noticeably as the number of samples increase, segment size should be limited based on the sampling rate and data type. For example, temperature measurements tend to change slower than voltage measurements, and a thermal fault typically takes longer to develop than a voltage anomaly does. Therefore, temperature measurement may need larger segments than voltage measurement.
\subsubsection{Cleaning} In this stage, invalid data points, like those that are out of the sensor's measurement range, are removed. However, there is no need to fill in the missing data due to this method's advantage.
\subsubsection{Preprocess} This stage includes filtering and data normalization. Segments where all signals do not show noticeable dynamics (changes) are excluded from the clustering process.
\subsubsection{K-shape clustering} In this stage, the K-shape algorithm given in Eq.~\ref{k-shape} is applied to the segment, during which all measurements are grouped into clusters based on the cross-similarity of their shapes. At the end of the clustering stage, each measurement is given its cluster membership $\mathbf{mem}[k]$ and the distance from its cluster's centroid $SBD(x_i,c_j)$.
\subsubsection{Anomaly Confirmation} As depicted in Fig.~\ref{fig_sim}, we look at two criteria during the anomaly confirmation stage. For the $kth$ segment, firstly, if one or multiple measurements changed its membership, it indicates an anomaly. Membership changes are detected by comparing the current membership $\mathbf{mem}[k]$ with the reference membership $\mathbf{mem}[0]$, which generates the anomaly indicator $a[k]$. Secondly, if there is no change in cluster membership ($a[k]=False$), we check if there is a significant increase in fitting errors/distance. To this end, each cluster's distance $dist[k]$ is compared to the reference $dist[0]$ and the previous cluster $dist[k-1]$. The first part is for capturing the accumulated changes associated with anomalies that developed gradually. E.g., thermal anomaly due to the battery's increased internal resistances. The second part is for capturing incremental changes associated with the anomaly that developed abruptly, such as a thermal anomaly caused by short-circuiting. The total distance function \cite{Tavenard2020TslearnData} is given by 
    \begin{equation} \label{dist_fcn}
        dist = \frac{1}{n}\sum_{i \in n}\left(\underset{j}{min}\left({SBD^2(x_i,c_j)}\right)\right)
    \end{equation}
where $SBD$ is given in Eq.~\ref{k-shape}. Afterwards, a weighting function is used to combined the two changes into a single output, defined as the anomaly confidence level $p[k]$. Details of the confirmation algorithm are given in Algorithm~\ref{alg}.

\section{Experimental results}\label{exp}

\subsection{Database and data sources}
The R\&D team at Gotion has built a cloud-based battery data platform. The data platform receives and cleans battery data from different sources such as fleet vehicles, onboard tests, and lab tests. Afterward, it is uploaded to a time-series database. There are also software modules for data visualization, battery simulation, and analysis. Most of the algorithms are implemented in Python. The proposed method is applied to battery data collected from fleet vehicle batteries between 2019 and 2021. The data sampling rate is about 0.1 Hz. There are different types of battery chemistry ($LiNiMnCo$ and $LiFePO_4$), pack configurations, and vehicle types.

\subsection{Initial results and discussions}

In this section, the testing results of two different battery packs are presented and compared to the fault detection based on the onboard BMS. The proposed algorithm is implemented in Python and deployed on a cloud computing platform (AWS). For the following testing results, the segment size is set to 25, equivalent to 4.1 min.  Notice that both cases have intermittent and longtime data losses. In the following figures, the data gaps are connected for better visualization.

\subsubsection{Case I}

In the first case, the proposed method is applied to a battery pack ($LiFePO_4$) that undergoes an over-temperature anomaly. As depicted in Fig.~\ref{fig2:1-1}, on Oct 30th, the battery's temperature near sensor \#13 rises significantly to over 70 \textcelsius. While both the proposed method and BMS can flag the anomaly, the new method is more than 90 min earlier. The detailed timing difference between the two methods is illustrated in Fig.~\ref{fig2:2-1}. As the figure shows, the BMS reports an over-temperature fault at 3:45 PM when the maximum temperature is over 55 \textcelsius. On the other hand, the proposed method sends an anomaly warning around 2:15 pm, just when sensor \#13 start to depart from the rest of the measurements. Fig.~\ref{fig2:2-2} shows a shape plot of the segment where the anomaly is detected. Clearly, one of the signal's (\#13) rising shape stands out from the rest. As signal \#13 continues to grow, its shapes for the following segments become less steep. Therefore, the confident level $d[k]$ is the largest at the beginning of the anomaly. This explains why the new method can send early warnings. In conclusion, case I validate that the proposed method can send an early warning for battery over-temperature faults.

\subsubsection{Case II}

\begin{figure}
    \centering
    \includegraphics[width=.5\textwidth]{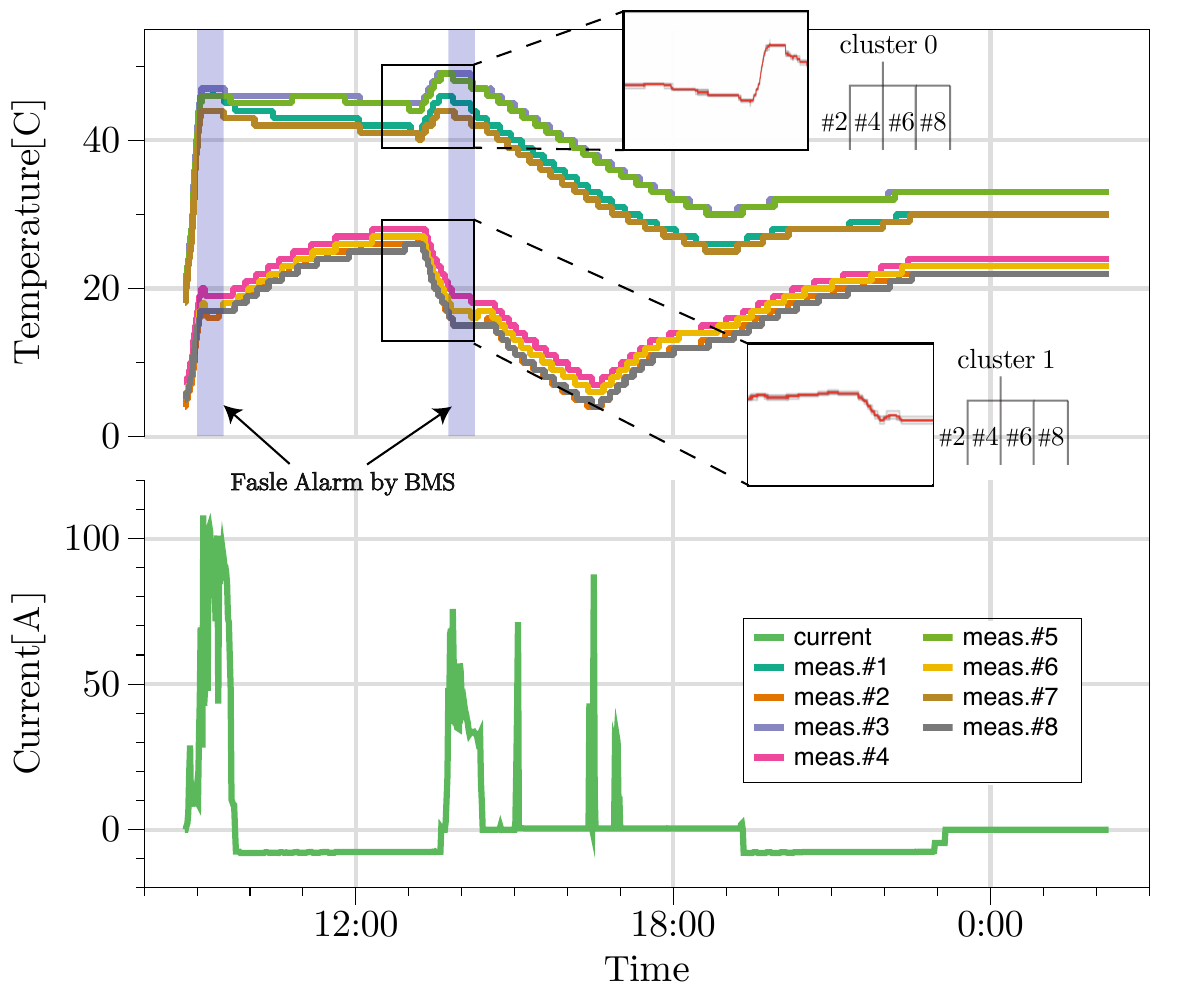}
    \caption{Case II: (a) Temperature measurements for a two-pack battery. (b) Current for the battery. Notice the temperature difference grows under large current loads.}  
    \label{fig:two_pack}
\end{figure}

The second case is from an EV with a two-pack configuration, in which two battery packs ($LiNiMnCo$) are installed in different locations inside the vehicle. As a result, the thermal measurements behave very differently. As Fig.~\ref{fig:two_pack} shows, the temperature difference between the two packs grows noticeably in 8:00-9:00 AM and 1:00-2:00 PM, when a large current is discharged from the battery. In both cases, the onboard BMS reports thermal fault despite that there is no real anomaly. The reason is that the onboard BMS uses hard thresholds from a look-up table. Also depicted in Fig.~\ref{fig:two_pack}, the proposed method successfully recognizes two clustering groups, including $\{1,3,5,7\}$ and $\{2,4,6,8\}$ and does not report any anomaly since there is no discrepancy found with the cluster groups. Note that the two zoom-in views in Fig.~\ref{fig:two_pack} are based on a longer segment for better illustration. In conclusion, this testing case shows that compared to BMS, the proposed method is more robust to variations caused by the pack design and sensor locations.

\begin{table}[hb]
    \caption{Description of the Vehicle Battery Data}
    \centering
    \begin{tabular}{lllll}
    \toprule
    \textit{field names} & \textit{data type} & \textit{value range} & \textit{unit} &\textit{source}\\
    \midrule
    cell voltage & double & $[0,5]$ & V & BMS\\
    temperature  & double & $[-50,100]$ & \textcelsius & BMS\\
    pack current & double & $[-500,500]$ & A & BMS\\
    battery faults & bool & $[0,1]$ & - & BMS\\
    battery status & enum & - & - & BMS\\
    \toprule
    \end{tabular}
    \label{sample-meta-data}
\end{table}

\section{Conclusion and Future work}
In this paper, we first identify common issues in the field of cloud-based battery anomaly/fault detection. Then, a method based on unsupervised shape-clustering is proposed for detecting battery thermal anomalies. The proposed method does not depend on a model or large training data. It also has several unique advantages, such as resilience to data loss and early detection capability. Finally, two test cases based on real vehicle data are studied. In one case, the new method is found to be more accurate than the BMS when applied to a multi-pack vehicle. In the other case study where a battery over-temperature fault occurs, the proposed method can flag the anomaly at the very early stage, more than 90 min ahead of the BMS.

\par

Future works include the following. Firstly, this method is currently being tested on large data sets. The rate of false positives and false negatives need to be investigated and compared to the BMS. Secondly, detection accuracy can be further improved by using other measurement signals like BMS status and battery currents.

\section{Acknowledgement}
We are thankful to our colleague, Asif Habeebullah, for his suggestions and insights.





\FloatBarrier
\bibliographystyle{IEEEtran}

\bibliography{references}

\end{document}